\documentclass[a4paper,fleqn,usenatbib]{mnras}

\usepackage{newtxtext,newtxmath}

\usepackage[T1]{fontenc}
\usepackage{ae,aecompl}

\usepackage{graphicx}	% Including figure files
\usepackage{amsmath}	% Advanced maths commands
\usepackage{xspace}
\usepackage{booktabs,caption}
\usepackage{threeparttable}
\usepackage{orcidlink}

\usepackage{color}
\usepackage{ulem}
\newcommand{\fdir}{./}
\newcommand{\bhb}{Gaia BH-like binary}
\newcommand{\paper}{paper}
\newcommand{\gaia}{{\it Gaia}}

\newcommand{\msun}{M_\odot}

\newcommand{\rsun}{R_\odot}

\title[Compact Binary Formation in Open Clusters]{Compact Binary
  Formation in Open Star Clusters I: High Formation Efficiency of
  Gaia BHs and Their Multiplicities}

\author[Ataru Tanikawa]{Ataru
  Tanikawa\orcidlink{0000-0002-8461-5517}$^{1}$\thanks{E-mail:
    tanikawa@ea.c.u-tokyo.ac.jp}, Savannah
  Cary\orcidlink{0000-0003-1860-1632}$^{2}$ , Minori
  Shikauchi\orcidlink{0000-0002-3561-8658}$^{3,4,5}$, Long
  Wang\orcidlink{0000-0001-8713-0366}$^{6,7}$, Michiko S.
  Fujii\orcidlink{0000-0002-6465-2978}$^{2}$ \\
$^{1}$Department of Earth Science and Astronomy, College of
  Arts and Sciences, The University of Tokyo, 3-8-1 Komaba, Meguro-ku,
  Tokyo 153-8902, Japan\\
$^{2}$Department of Astronomy, Graduate School of Science, The
  University of Tokyo, 7-3-1 Hongo, Bunkyo-ku, Tokyo 113-0033, Japan\\
$^{3}$Department of Physics, the University of Tokyo, 7-3-1 Hongo,
  Bunkyo, Tokyo 113-0033, Japan \\
$^{4}$Research Center for the Early Universe (RESCEU), the University
  of Tokyo, 7-3-1 Hongo, Bunkyo, Tokyo 113-0033, Japan \\
$^{5}$Department of Physics and Astronomy, the University of British
  Columbia, 6224 Agricultural Road, Vancouver, BC, V6T 1Z1, Canada\\
$^{6}$School of Physics and Astronomy, Sun Yat-sen University, Daxue
  Road, Zhuhai, 519082, China\\
$^{7}$CSST Science Center for the Guangdong-Hong Kong-Macau Greater
  Bay Area, Zhuhai, 519082, China\\}

\date{Accepted XXX. Received YYY; in original form ZZZ}

\pubyear{2017}

\begin{document}
\label{firstpage}
\pagerange{\pageref{firstpage}--\pageref{lastpage}}
\maketitle

\begin{abstract}

  Gaia BHs, black hole (BH) binaries discovered from database of an
  astrometric telescope \gaia, pose a question to the standard binary
  evolution model. We have assessed if Gaia BHs can be formed through
  dynamical capture in open clusters rather than through isolated
  binary evolution. We have performed gravitational $N$-body
  simulations of $100$ open clusters with $10^5 \msun$ in total for
  each metallicity $Z=0.02$, $0.01$, and $0.005$. We have discovered
  one \bhb\ escaping from an open cluster, and found that the
  formation efficiency of Gaia BHs in open clusters ($\sim 10^{-5}
  \msun^{-1}$) is larger than in isolated binaries ($\sim 10^{-8}
  \msun^{-1}$) by 3 orders of magnitude. The \bhb\ is the inner binary
  of a triple star system. Gaia BHs can have tertiary stars
  frequently, if they are formed in open clusters. Combining
  additional $N$-body simulations with 8000 open clusters with $8
  \times 10^6 \msun$, we have estimated the number of Gaia BHs in the
  Milky Way disk to $10^4 - 10^5$ (depending on the definitions of
  Gaia BHs), large enough for the number of Gaia BHs discovered so
  far. Our results indicate that the discoveries of Gaia BHs do not
  request the reconstruction of the standard binary evolution model,
  and that Gaia BHs are a probe for the dynamics of open clusters
  already evaporated.
  
\end{abstract}


\section{Introduction}
\label{sec:Introduction}

Black holes (BHs) are the final state of massive stars. Albeit of
their darkness, they have been discovered in binary systems, such as
X-ray binaries \citep{2017hsn..book.1499C}, spectroscopic binaries
\citep{2022NatAs...6.1085S}, astrometric binaries
\citep{2023MNRAS.518.1057E, 2023MNRAS.521.4323E, 2023ApJ...946...79T,
  2023AJ....166....6C}, and gravitational wave sources
\citep{2021arXiv211103606T}. The discoveries of astrometric BH
binaries, also known as Gaia BH1 and BH2
\citep[][respectively]{2023MNRAS.518.1057E, 2023MNRAS.521.4323E} in
Gaia DR3 \citep{2022arXiv220605595G}, challenge the standard binary
evolution model. Gaia BH1 and BH2 (hereafter, Gaia BHs) have $\sim 1
\msun$ visible stars as companions, orbital periods of $\sim
10^2$-$10^3$ days, and moderately high eccentricities ($\sim
0.5$). Isolated binaries can form such configurations only if common
envelope ejection is $\sim 10$ times more efficient than previously
expected \citep{2023MNRAS.518.1057E}.

Another possible formation channel of Gaia BHs is that a BH
dynamically captures a $\sim 1 \msun$ visible star in open
clusters. This is similar to double BH formation in open clusters
\citep{2014MNRAS.441.3703Z, 2019MNRAS.486.3942K, 2020MNRAS.495.4268K,
  2019MNRAS.487.2947D, 2021MNRAS.503.3371B, 2021MNRAS.507.3612R}. As
for double BHs, such dynamical capture can happen in globular clusters
\citep{2000ApJ...528L..17P, 2013MNRAS.435.1358T, 2016PhRvD..93h4029R,
  2017MNRAS.464L..36A, 2022ApJ...934L...1K}, first star clusters
\citep{2022MNRAS.515.5106W}, and galactic centers
\citep{2020ApJ...898...25T}.  However, Gaia BHs should not originate
from globular clusters, first star clusters, and galactic centers,
since they are Milky Way disk components, and have near solar
metallicities. They are likely to be formed in open clusters if their
formation channel is dynamical capture.

Gaia BH1 and BH2 have different orbital periods by an order of
magnitude, and different types of visible stars: a main-sequence (MS)
star for Gaia BH1 and a red giant star for Gaia BH2. They may have
different origins. What can be said with certainty is that Gaia BHs
have common features in which they are difficult to be formed through
isolated binary evolution, if common envelope is ejected efficiently
\citep{2023MNRAS.518.1057E, 2023MNRAS.521.4323E}. Thus, we explore
possibility that Gaia BHs are formed through dynamical capture in open
clusters.

In this \paper, we report an extremely high formation efficiency of
Gaia BHs in open clusters. Moreover, our simulations show that they
frequently host tertiary stars.  The structure of this \paper\ is as
follows. In section \ref{sec:Method}, we describe our numerical method
to follow Gaia BH formation in open clusters and isolated binaries. In
section \ref{sec:Results}, we show our simulation results. In section
\ref{sec:ConclusionAndDiscussion}, we conclude our results, and
discuss the implication of our results for Gaia BHs discovered
recently. We focus on a \bhb\ formed in 100 open clusters with $10^5
\msun$ and metallicity of $Z=0.005$, while we discuss the dependence
of the formation efficiency of Gaia BHs on initial conditions of open
clusters in Appendix.

\section{Method}
\label{sec:Method}

We follow the dynamical evolution of open clusters by means of a
gravitational $N$-body code {\tt PeTar}
\citep{2020MNRAS.497..536W}. The {\tt PeTar} code is based on the
particle-tree and particle-particle algorithm
\citep{2011PASJ...63..881O, 2017PASJ...69...81I} with the aid of {\tt
  FDPS} \citep{2016PASJ...68...54I, 2020PASJ...72...13I}. Binary
orbital evolution and close encounters are treated by the slow-down
algorithmic regularization method \citep[{\tt
    SDAR}:][]{2020MNRAS.493.3398W}. The external potential is
calculated by {\tt GALPY} \citep{2015ApJS..216...29B}. We can see
comparison between {\tt PeTar} and {\tt NBODY6++GPU}
\citep{2012MNRAS.424..545N, 2015MNRAS.450.4070W, 2022MNRAS.511.4060K}
for mass scale of small globular clusters in
\cite{2020MNRAS.497..536W}, and open clusters in Appendix
\ref{sec:ComparisonBetweenPetarAndNbody6}.

The {\tt PeTar} code is coupled with the {\tt BSE} code
\citep{2002MNRAS.329..897H, 2020A&A...639A..41B} to solve single and
binary star evolutions.  We overview the model of single and binary
star evolution. Single stars evolve on a track developed by
\cite{2000MNRAS.315..543H} with stellar winds formulated by
\cite{2010ApJ...714.1217B}. Massive stars either cause supernovae or
collapse to BHs at the end of their lives, and leave behind NSs and
BHs with masses modeled by the rapid model \citep{2012ApJ...749...91F}
with modification of pair instability supernovae modeled by
\cite{2016A&A...594A..97B}. Binary evolution model is the same as
\cite{2020A&A...639A..41B}. It contains various binary evolution
processes, such as tidal interaction, wind accretion, stable mass
transfer, and common envelope. Here, we only describe parameters of
common envelope which Gaia BHs formation is the most sensitive to
\citep{2023MNRAS.518.1057E}. A binary can experience common envelope
evolution modeled as the $\alpha$ formalism
\citep{1984ApJ...277..355W}, where we adopt $\alpha=3$ and $\lambda$
of \cite{2014A&A...563A..83C}.

We generate $100$ open clusters for each metallicity of $Z=0.02$,
$0.01$, and $0.005$. The total mass of each open cluster is $10^3
\msun$, and then the total mass of open clusters is $10^5 \msun$ for
each metallicity.  We adopt the Plummer model for the initial phase
space distribution of stars. We set the half-mass radius to $0.8$
pc. Each cluster orbits around Milky Way at $8$ kpc with $220$ km
s$^{-1}$. The initial binary fraction is $100$ \%. Although the binary
fraction appears too high, it is not sensitive to the formation
efficiency and multiplicity of Gaia BHs as seen in Appendix
\ref{sec:EffectsOfInitialBinaryFraction}. The primary stars of
binaries have Kroupa's initial mass function
\citep{2001MNRAS.322..231K} with the minimum and maximum masses of
$0.08$ and $150 \msun$, respectively. The mass ratio, orbital period,
and eccentricity distribution of binaries follow initial conditions
from \cite{2012Sci...337..444S}. The initial conditions are generated
using {\tt MCLUSTER} \citep{2011MNRAS.417.2300K}. We follow their
evolutions over $500$ Myr. Although the initial central number density
seems to be high ($\sim 10^3$ pc$^{-3}$), the central number density
of visible stars (defined as stars except BHs, neutron stars, and
white dwarfs) drop to $\sim 3$ pc$^{-3}$ at $500$ Myr.

For comparison, we also calculated isolated binary evolution.  We
prepare two types of initial conditions for calculations of isolated
binary evolution. Like the binaries in our open clusters, the 1st type
of initial conditions for isolated binary evolution are also adopted
from \cite{2012Sci...337..444S}. We find that the 1st type of initial
conditions never form Gaia BHs, because the minimum mass ratio is
$0.1$. In our single star evolution model, $\gtrsim 20 \msun$ stars
can leave behind BHs. BHs cannot have $\lesssim 1 \msun$ stars as
companions. Thus, we prepare the 2nd type of initial conditions in
which the minimum mass ratio is reduced to $0.0005$, such that even
$150 \msun$ stars can have $0.08 \msun$ stars as companions. We would
like to remark that the 2nd type of initial conditions is much more
likely to form Gaia BHs than the initial conditions of open clusters
unless anything prohibits Gaia BHs from forming. We also note that we
do not adopt such a small mass ratio for initial binaries in open
clusters. We follow their evolution over $30$ Myr. Until that, all BH
progenitors evolve to BHs.

Throughout this \paper, we compare properties of BH binaries for open
clusters at $500$ Myr, and for isolated binaries at $30$ Myr. We
underestimate the formation efficiency of BH binaries with secondary
masses of $\gtrsim 2.5 \msun$, because $\gtrsim 2.5 \msun$ stars
evolve to remnant objects until $500$ Myr. However, this does not
affect our purpose. Gaia BHs discovered so far have secondary stars of
$\sim 1 \msun$.

Our simulations generate $59$ BH binaries. Such BH binaries have
various secondary masses, periods, and eccentricities. In this \paper,
we define Gaia BHs as BH binaries with secondary masses of $\le 1.1
\msun$, periods of $100$-$2000$ days, and eccentricities of
$0.3$-$0.9$ unless stated.

\section{Results}
\label{sec:Results}

We find that one open cluster with $Z=0.005$ contains a BH binary
similar to Gaia BHs, hereafter called ``\bhb''. We summarize the
properties of the \bhb\ in Table \ref{tab:SummaryOfBhbinary}. Its BH
mass is $21.4 \msun$. Its companion (hereafter, secondary) star is a
MS star and has a mass of $0.82 \msun$. The binary period is $8.3
\times 10^2$ days, and its eccentricity oscillates from $0.3$ to $0.8$
due to von Zeipel-Lidov-Kozai (ZLK) mechanism
\citep{1910AN....183..345V, 1962P&SS....9..719L,
  1962AJ.....67..591K}. We find that the \bhb\ is the inner binary of
a triple star system, and it is because of a tertiary star that we see
ZLK oscillation of the \bhb.  The tertiary star is a MS star and has
$1.59 \msun$ at $500$ Myr. The outer period and eccentricity are $1.2
\times 10^6$ days and $0.689$, respectively.

\begin{table}
  \centering
  \caption{Summary of the \bhb at $500$
    Myr.} \label{tab:SummaryOfBhbinary}
  \begin{tabular}{l|ll}
    \hline
    Parameters      & Values & Remarks \\
    \hline
    \hline
    Metallicity (Z) & $0.005$ \\
    \hline
    BH mass         & $21.4 \msun$ \\
    Secondary mass  & $0.82 \msun$           & MS star \\
    Period          & $8.3 \times 10^2$ days \\
    Eccentricity    & $0.3$-$0.8$            & Oscillating \\
    \hline
    Tertiary mass       & $1.59 \msun$           & MS star \\
    Outer period        & $1.2 \times 10^6$ days \\
    Outer eccentricity  & $0.689$ \\
    Mutual inclination  & $34$-$59$ deg           & Oscillating \\
    \hline
  \end{tabular}
\end{table}

\begin{figure}
  \includegraphics[width=\columnwidth]{\fdir/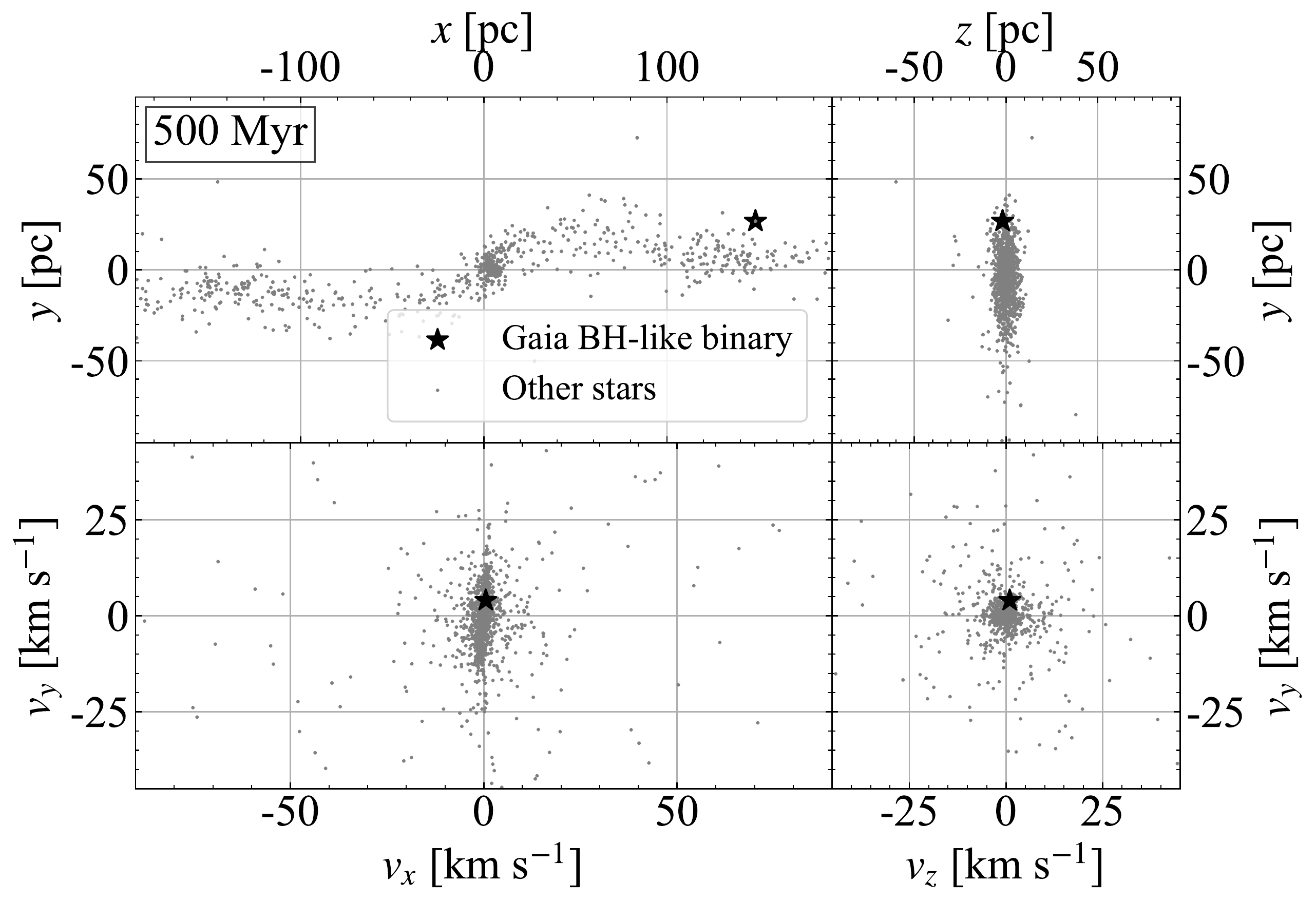}
  \caption{Positions (top) and velocities (bottom) of the \bhb\ (star
    points) and other stars (dots) at $500$ Myr. All the stars
    including BHs, neutron stars, and white dwarfs, are plotted. The
    central number density of visible stars is $\sim 3.4$ pc$^{-3}$.}
  \label{fig:phasespace}
\end{figure}

Figure \ref{fig:phasespace} shows the position and velocity of the
\bhb\ at $500$ Myr. We can see that the \bhb\ has escaped from the
open cluster located at the coordinate origin. The \bhb\ will be
long-lived after $500$ Myr as long as it is not perturbed by the
tertiary star (described later). This \bhb\ escapes from the open
cluster through either of two-body relaxation or close binary
interaction\footnote{Such a binary with massive objects can be ejected
from an open cluster \citep{2011Sci...334.1380F}}. Looking at their
position and velocity in Figure \ref{fig:phasespace}, the
\bhb\ belongs to a tidal tail of the open cluster. This means that it
will be identified as a Galactic disk component long after the open
cluster has evaporated.

\begin{figure}
  \includegraphics[width=\columnwidth]{\fdir/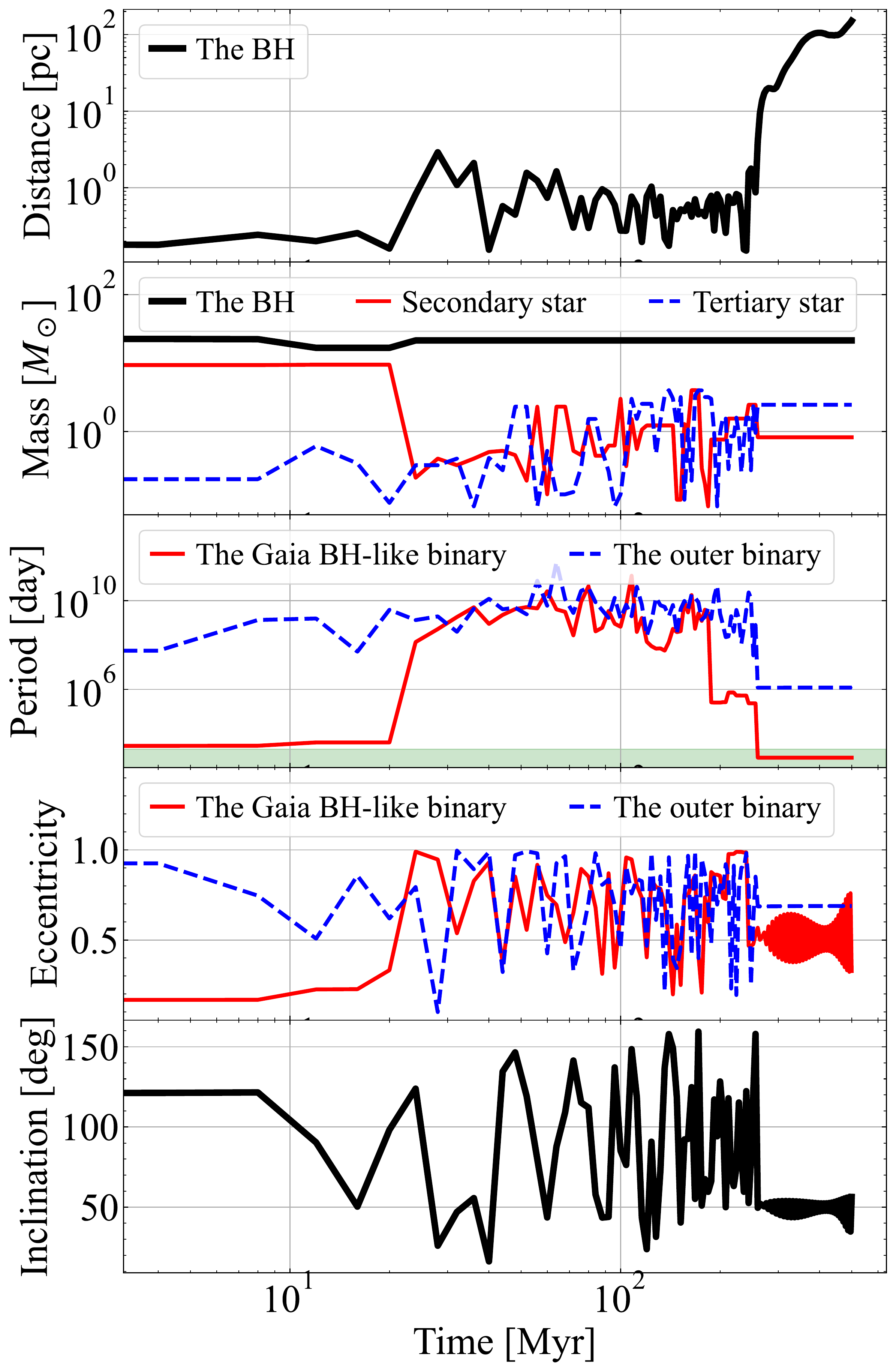}
  \caption{Time evolution of the \bhb\ and its progenitor. Secondary
    and tertiary stars are defined as temporary companions of the BH
    and its progenitor at each time.  We indicate the distance from
    the center, component masses, binary periods, binary
    eccentricities, and mutual inclination from top to bottom. These
    quantities are plotted for each $4$ Myr. The shaded regions show
    binary periods of $100-2000$ days, which tends to be detected by
    \gaia.}
  \label{fig:history}
\end{figure}

\begin{figure}
  \includegraphics[width=\columnwidth]{\fdir/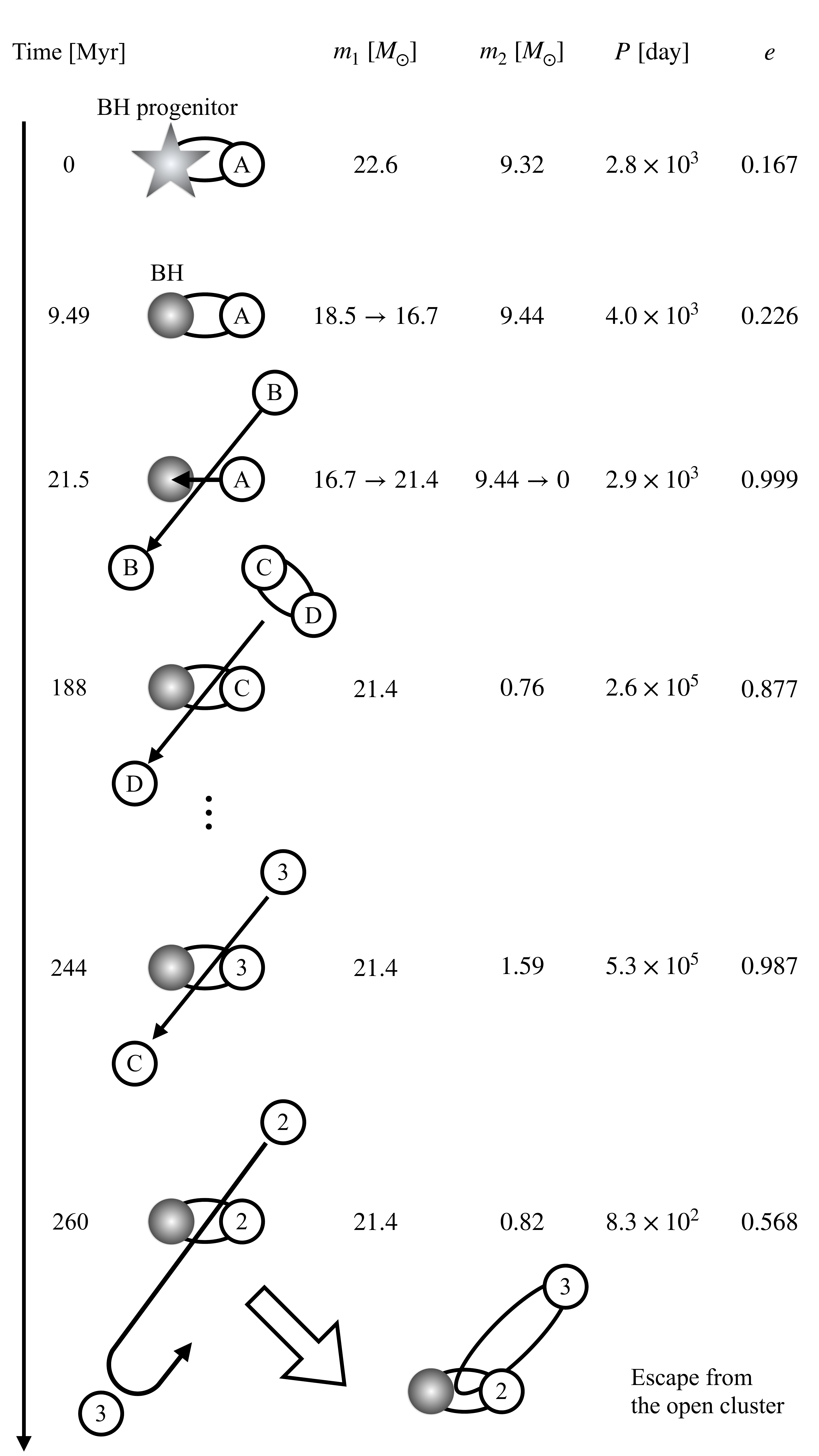}
  \caption{Illustration of the \bhb\ formation. Objects indicated by
    ``BH'', ``2'' and ``3'' are the BH, secondary and tertiary stars
    in the \bhb, respectively. Object ``A'' is the initial companion
    of the BH progenitor. Objects ``B'', ``C'', ``D'' are perturbers
    for the BH and its companions. Although the \bhb\ experience many
    interactions during $188$-$244$ Myr, we omit them. Thus, an
    ejected object at $244$ Myr is not actually object ``C''.}
  \label{fig:picture}
\end{figure}

We show the time evolution of the \bhb\ and its progenitor in Figure
\ref{fig:history}, and illustrate the formation process of the
\bhb\ in Figure \ref{fig:picture}. At the initial time, its BH
progenitor has $22.6 \msun$ and a companion with $9.32 \msun$. The
binary period and eccentricity are $2.8 \times 10^3$ days and $0.167$,
respectively. At $9.49$ Myr, its BH progenitor evolves to a $16.7
\msun$ BH. At $21.5$ Myr, this binary is perturbed by an open cluster
star, excites its eccentricity to nearly 1, and merges. This merger
leaves behind a $21.4 \msun$ BH. Until $188$ Myr, this BH does not
form any hard binary. Although soft binaries are occasionally formed,
their binary periods are at least $\sim 10^7$ days. Such soft binaries
are quickly disrupted by perturbations of other stars. At $188$ Myr,
the BH captures an open cluster star, and form a relatively hard
binary with a period of $5.3 \times 10^5$ days. After this, the BH
companion is replaced with another star several times (although
omitted in Figure \ref{fig:picture}). At $244$ Myr the BH companion
becomes a $1.59 \msun $ MS star to be finally the tertiary star. At
$260$ Myr, the BH companion is exchanged with a $0.82 \msun$ MS star
to be finally the secondary star. The superseded companion is not
fully ejected from the \bhb, and stays as the tertiary star. Soon
after this interaction, this \bhb\ with the tertiary star escapes from
the open cluster.

We can see in Figure \ref{fig:history} that the \bhb\ eccentricity and
mutual inclination between the orbital planes of the inner and outer
binaries (hereafter, mutual inclination) are oscillating after the
\bhb\ escapes due to ZLK mechanism. The ZLK oscillation is modulated,
and the modulation amplitude appears inconstant. This is because the
cadence of the snapshot ($4$ Myr) is longer than the timescale of the
ZLK oscillation. In order to catch the ZLK oscillation by sufficiently
high cadence of the snapshot, we recalculate the orbital evolution of
the triple system. For this purpose, we extract the masses, positions,
and velocities of the triple-system components at $400$ Myr, and
follow their evolution over $2$ Gyr by means of the {\tt SDAR} code
\citep{2020MNRAS.493.3398W}, where the SDAR code is incorporated into
the {\tt PeTar} code to treat close binary and multiple star systems
as described above. We approximate that the triple system is
completely isolated from other stars, since the triple system escapes
from the open cluster at $400$ Myr. We do not take into account the
tertiary star evolution to a red giant and white dwarf phases.

\begin{figure}
  \includegraphics[width=\columnwidth]{\fdir/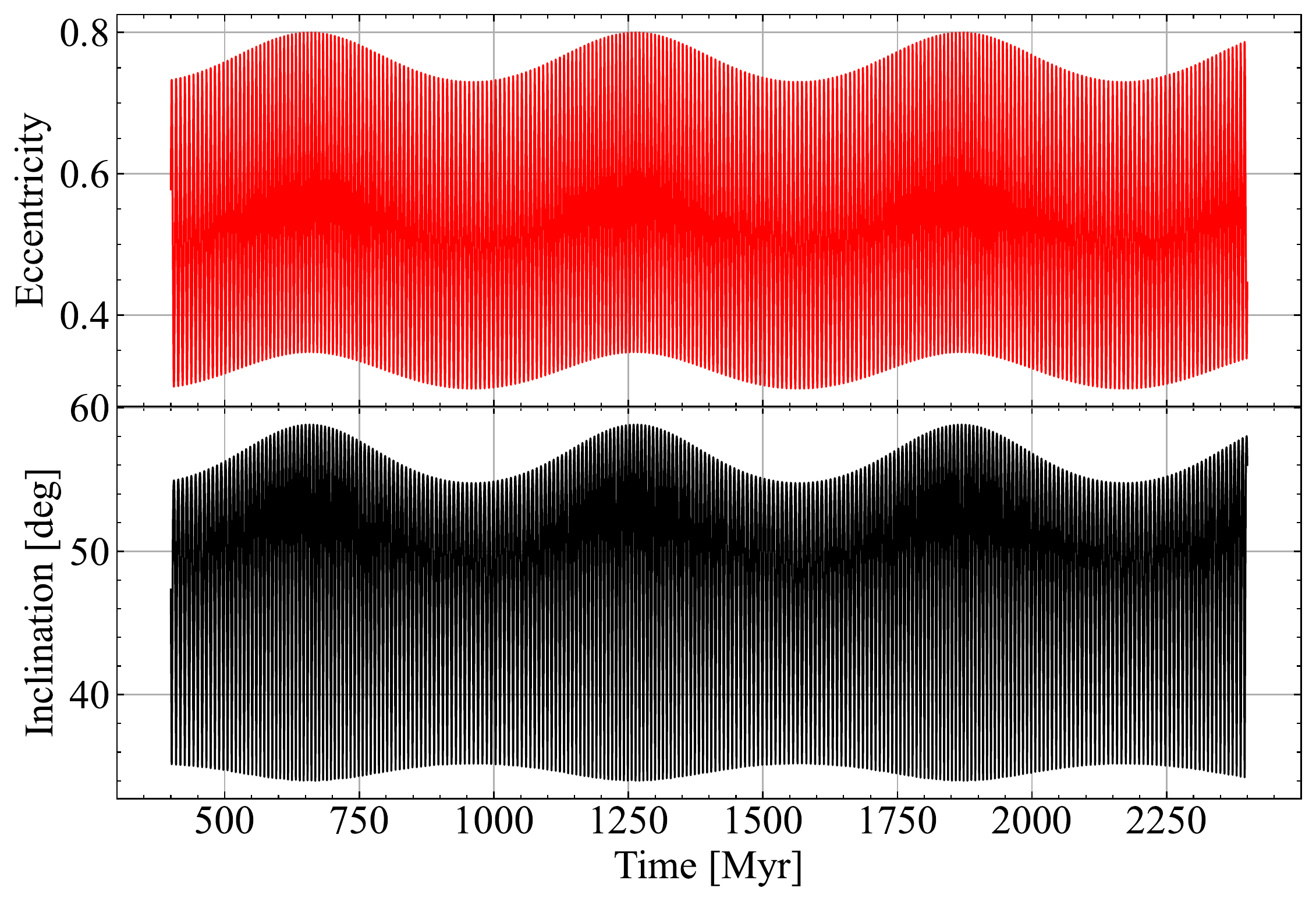}
  \caption{Time evolution of the \bhb\ eccentricity (top) and mutual
    inclination (bottom) over $2$ Gyr from $400$ Myr. The quantities
    are plotted for each $0.1$ Myr.}
  \label{fig:tripleHighCadence}
\end{figure}

Figure \ref{fig:tripleHighCadence} shows the eccentricity and mutual
inclination of the \bhb. The cadence of the snapshot is $0.1$ Myr,
which is sufficiently high to show the ZLK oscillation. The
oscillation on several Myr and modulation on several $100$ Myr are due
to the quadrupole-level and octupole-level interactions, respectively
\citep{2015MNRAS.452.3610A}. The modulation amplitude is constant,
although it appears not to be constant due to the low cadence of the
snapshot ($4$ Myr) in Figure \ref{fig:history}. We can see that the
\bhb\ eccentricity rises to at most $0.8$, and its pericenter distance
is reduced down to at least $0.97$ au. Since this pericenter distance
is much larger than the secondary radius, the \bhb\ will not
merge. Note that the secondary star, a $0.82$ MS star, will not evolve
to a giant star within the Hubble time. The tertiary star evolves to a
red giant star and then a white dwarf over $\sim 2$ Gyr. The red giant
star will not fill the Roche lobe of the outer binary, because the
pericenter distance of the outer binary is $2.0 \times 10^2$ au much
larger than the asymptotic giant branch radius $\sim 1$ au. The
tertiary star gradually decreases its mass and influence on the
\bhb. The tertiary star evolution is expected not to trigger merger or
disruption of the \bhb.

\begin{figure}
  \includegraphics[width=\columnwidth]{\fdir/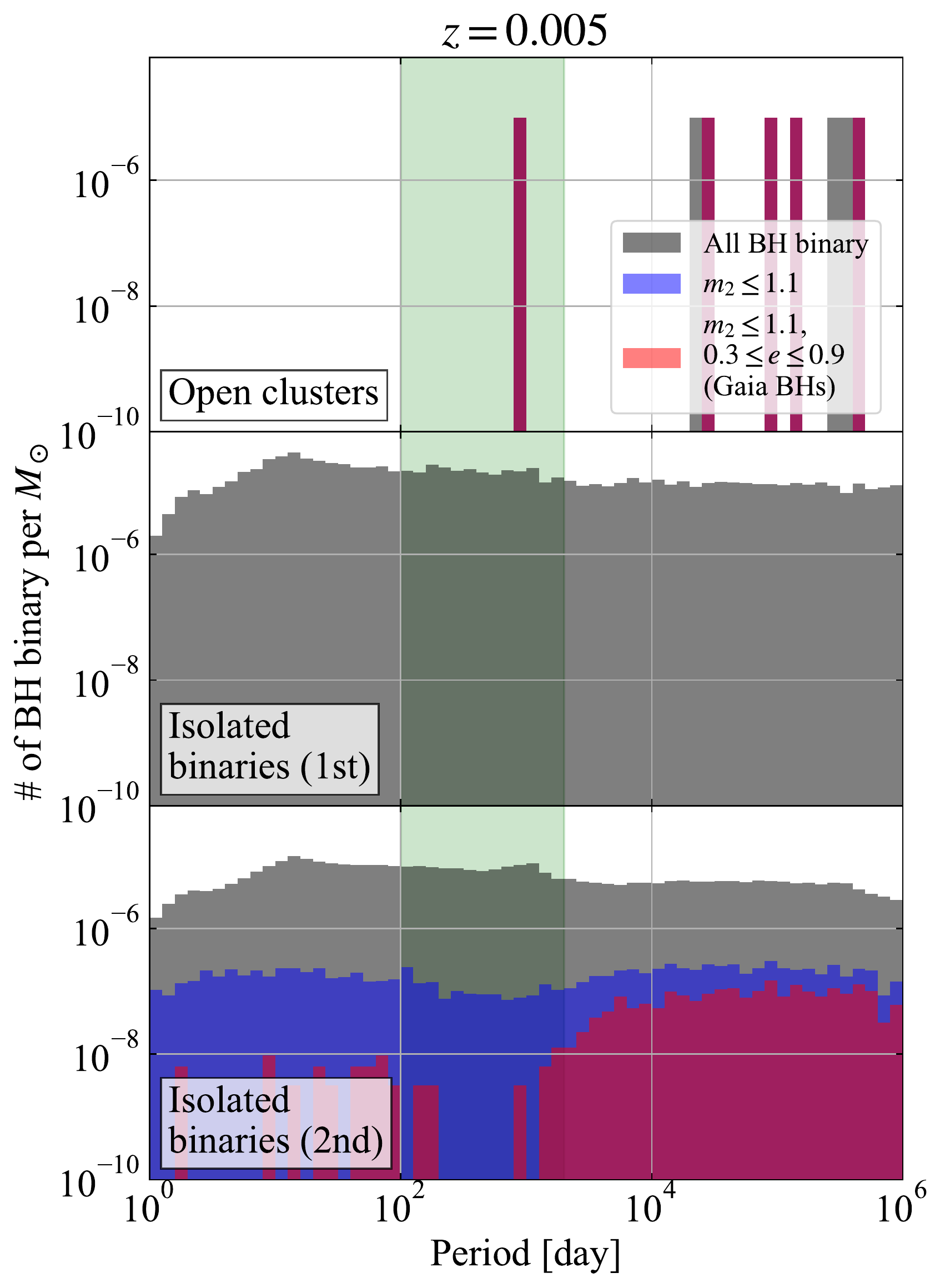}
  \caption{Formation efficiency of BH binaries per $\msun$ for each
    bin in open clusters (top), the 1st type of isolated binaries
    (middle), and the 2nd type of isolated binaries (bottom). The
    stellar type of the companions are limited to MS stars. The grey,
    blue, and red histograms indicate All BH binaries, BH binaries
    with $\le 1.1 \msun$ companions, and those with $\le 1.1 \msun$
    companions and eccentricities from $0.3$ to $0.9$ (i.e. Gaia
    BHs). The shaded regions show binary periods of $100-2000$ days,
    which tends to be detected by \gaia.}
  \label{fig:period_kw14_z5e-03}
\end{figure}

In Figure \ref{fig:period_kw14_z5e-03}, we compare the formation
efficiency among BH binaries in open clusters to the 1st and 2nd types
of isolated binary evolution. We focus on only binary periods of
$100-2000$ days, which tends to be detected by \gaia\footnote{The
binary periods of Gaia BH1 and BH2 are $\sim 180$ day
\citep{2023MNRAS.518.1057E} and $\sim 1300$ day
\citep{2023MNRAS.521.4323E, 2023ApJ...946...79T}.}. The formation
efficiency of all BH binaries in open clusters is much smaller than
both the types of isolated binaries. However, when we limit BH
binaries to Gaia BHs, i.e. those with light companions ($1.1 \msun$)
and moderately high eccentricities ($0.3-0.9$), the formation
efficiency is reversed. It is $\sim 10^{-5} \msun^{-1}$ in open
clusters, while it is $\sim 10^{-8} \msun^{-1}$ even in the 2nd type
of isolated binaries. Thus, the formation efficiency is larger by $3$
orders of magnitudes.

We again emphasize that the formation of Gaia BHs in open clusters is
much higher than in isolated binaries with the respect of initial
conditions. The numbers of binaries with $>20 \msun$ and $<1.1 \msun$
stars are $0$, $0$, and $5.4 \times 10^{-5}$ per $1 \msun$ in open
clusters and the 1st and 2nd types of isolated binaries,
respectively. If we limit their orbital periods to $100-2000$ days,
the number is $9.2 \times 10^{-6}$ per $1 \msun$ in the 2nd types of
isolated binaries. Note that $>20 \msun$ stars evolve to BHs long
before $500$ Myr, and $<1.1 \msun$ stars are still MS stars at $500$
Myr if they evolve as single stars; such binaries can be potentially
progenitors of Gaia BHs. On the other hand, the formation efficiencies
of Gaia BHs are $\sim 10^{-5}$, $0$, and $\sim 10^{-8} \msun^{-1}$ in
open clusters and the 1st and 2nd types of isolated binaries,
respectively. Open clusters are more disadvantageous to form Gaia BHs
than the 2nd type of isolated binaries. However, open clusters can
form many more Gaia BHs.

\begin{figure}
  \includegraphics[width=\columnwidth]{\fdir/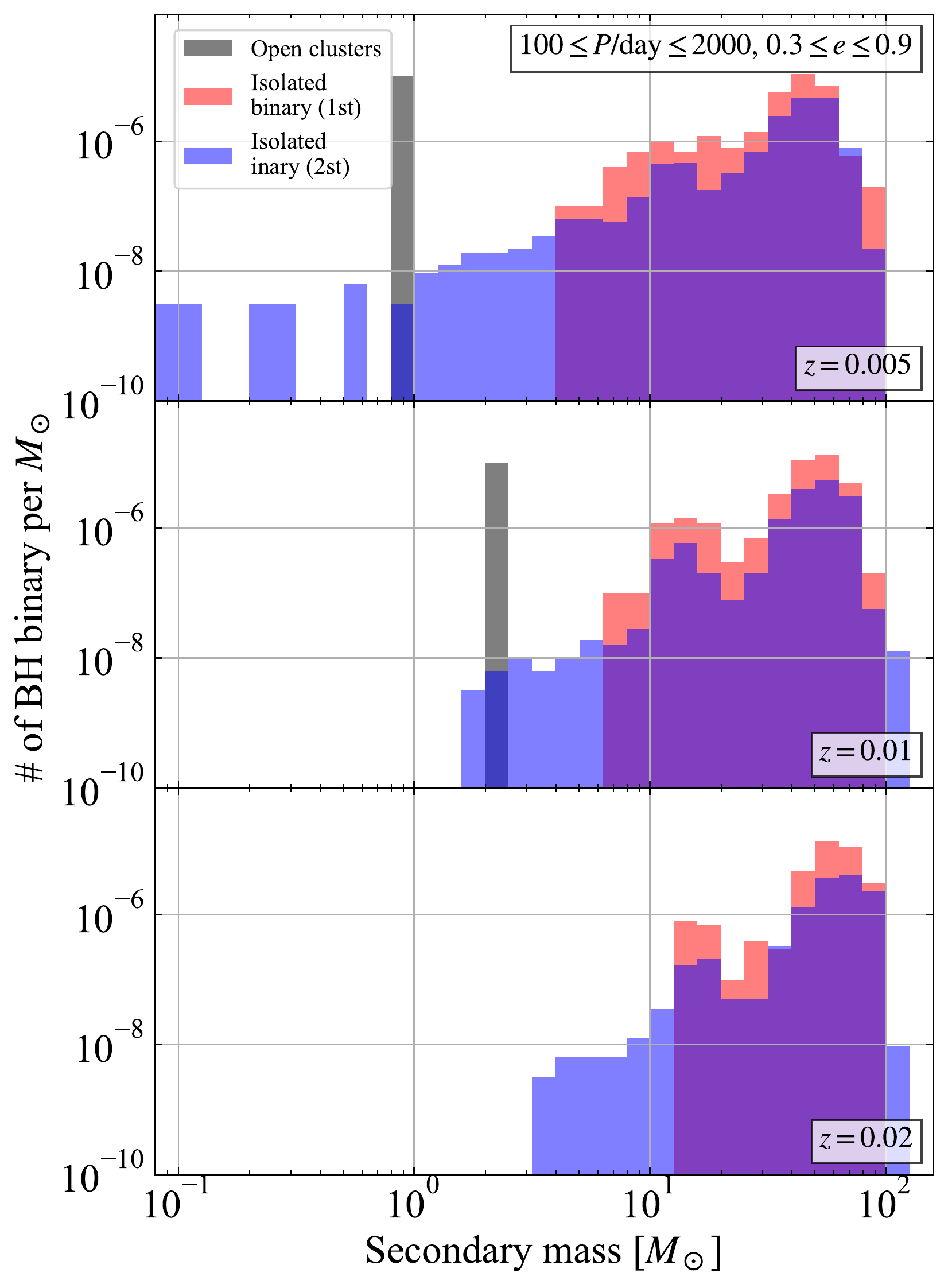}
  \caption{Formation efficiency of BH binaries per $\msun$ for each
    bin in open clusters, and the 1st and 2nd types of isolated
    binaries as a function of secondary masses. The top, middle, and
    bottom panels indicate $Z=0.005$, $0.01$, and $0.02$,
    respectively. All the secondary stars are MS stars, and all the
    binary periods and eccentricities are $100$-$2000$ days and
    $0.3$-$0.9$, respectively.}
  \label{fig:secondarymass_kw14}
\end{figure}

Figure \ref{fig:secondarymass_kw14} shows the formation efficiency of
BH binaries as a function of secondary masses. From the top panel, we
can confirm that, for the secondary mass of less than $1.1 \msun$, the
formation efficiency in open clusters is larger than in isolated
binaries by 3 orders of magnitude for $Z=0.005$. If BH binaries with
the secondary mass of $\sim 2 \msun$ can be regarded as Gaia BHs,
there is another Gaia BH in a $Z=0.01$ open cluster (see the middle
panel in Figure \ref{fig:secondarymass_kw14}). Such a BH binary is
also formed in open clusters more easily than in isolated binaries by
3 orders of magnitude. Note that we find no tertiary star for this BH
binary.

The mass distribution in Figure \ref{fig:secondarymass_kw14} may have
the sharp peak at $\sim 1 \msun$ for the open clusters with $Z=0.005$
due to the small sampling. If we increase the number of open cluster
simulations to much more than $100$, the sharp peak is likely to be
smeared out around $\sim 1 \msun$. Thus, some of them may exceed $1
\msun$, and may not be categorized as Gaia BHs. Nevertheless, it
should have a minor effect on the formation efficiency of Gaia BHs in
open clusters.

The low formation efficiency in isolated binaries is consistent with
the argument of \cite{2023MNRAS.518.1057E}. Here, we review its
reason. Let's consider a binary with a BH progenitor ($\gtrsim 20
\msun$) and a $\sim 1 \msun$ star. Its period is $\sim 10^3$ days, and
then its separation is $\sim 10^3 \rsun$. When the BH progenitor
evolves to a giant star, it fills its Roche lobe, and starts Roche
lobe overflow. The Roche lobe overflow is typically unstable, since
the donor star (here, the BH progenitor) is much more massive than the
accretor (here, the $1 \msun$ star). Thus, the binary experiences
common envelope evolution. The common envelope evolution ends with a
merger of the BH progenitor core and $\sim 1 \msun$ star. This is
because the gravitational binding energy between the core and star is
too small to fully eject the common envelope. Even if the binary
survives against the common envelope evolution, its orbital period is
much shorter than the orbital periods of Gaia BHs, and its
eccentricity is reduced to $\lesssim 0.2$
\citep[e.g.][]{2021ApJ...920...86K, 2022PhRvD.106d3014T}. The BH natal
kick could excite its eccentricity if its velocity is sufficiently
high. However, the barycenter velocities of the Gaia BHs are too small
to suppose such high BH natal kick velocities.

In contrast to isolated binaries, the formation of Gaia BHs in open
clusters avoids all the difficulties described above. As illustrated
in Figure \ref{fig:picture}, Gaia BHs can be formed through dynamical
capture after BH progenitors evolve to BHs in open clusters. If we
dare to mention the difficulty of Gaia BH formation in open clusters,
BHs may rarely capture $\sim 1 \msun$ stars. Generally, BHs capture
more massive stars more easily, and there are many $\gtrsim 1 \msun$
MS and white dwarfs in open clusters.

Taking into account only the \bhb, we estimate the number of Gaia BHs
formed in open clusters over Milky Way as follows:
\begin{align}
  N_{\rm Gaia BH,MW} &\sim 1.6 \times 10^4 \left( \frac{\eta}{10^{-5}
    \msun} \right) \left( \frac{M_{\rm MW}}{6.1 \times 10^{10} \msun}
  \right) \nonumber \\
  &\times \left( \frac{f_{Z=0.005}}{0.26} \right) \left( \frac{f_{\rm
      cluster}}{0.1} \right), \label{eq:GaiaBH1}
\end{align}
where $\eta$ is the formation efficiency obtained from our simulation
results, $M_{\rm MW}$ is Milky Way stellar mass
\citep[e.g.][]{2015ApJ...806...96L}, and $f_{Z=0.005}$ is the mass
fraction of $Z=0.005$ in the Milky Way, which is calculated by
integrating stellar mass between $Z=0.003$ and $0.007$ in a Milky Way
model of \cite{2022ApJ...937..118W} and
\cite{2023arXiv230107207S}. The fraction $f_{\rm cluster}$ is the mass
fraction of stars formed in open clusters \citep{2006A&A...459..113M,
  2007A&A...468..151P}. We assume that the lifetime of Gaia BHs is
larger than the Hubble time because of the small mass of the \bhb.
Since the size of Milky Way disk is $\sim 10$ kpc, the number of Gaia
BHs can be $\sim 160$ within $\sim 1$ kpc. We expect that the
formation efficiency in open clusters will explain the number of Gaia
BHs, even if the number of Gaia BHs (currently, $2$) will grow in the
future as seen in the lists of BH binary candidates
\citep{2022arXiv220700680A, 2023MNRAS.518.2991S}.

\section{Conclusion and Discussion}
\label{sec:ConclusionAndDiscussion}

Our main conclusions in this \paper\ are as follows.
\begin{enumerate}
\item Formation efficiency of Gaia BHs in open clusters is higher than
  in isolated binaries by $3$ orders of magnitude.
\item Gaia BHs are in multiple star systems with a high probability.
\end{enumerate}

We find a \bhb\ in our simulations albeit of the small total mass of
open clusters. The \bhb\ has many points similar to Gaia BHs. The
visible star has $\sim 1 \msun$, and the binary period and
eccentricity are $10^2$-$10^3$ days and $\sim 0.5$, respectively. Its
barycenter orbit indicates that it is a Galactic disk component. Since
the visible star is dynamically captured by the BH after the BH
formation, it should not be polluted by supernova ejecta from the BH,
which is consistent with the chemical abundance pattern of Gaia BH2
\citep{2023MNRAS.521.4323E}. The visible star has subsolar metallicity
similar to Gaia BH1, whose visible star has ${\rm [Fe/H]} =-0.2 \pm
0.05$ \citep{2023MNRAS.518.1057E}. The number of such BH binaries is
large enough to explain the presence of Gaia BHs. The \bhb\ contains
different points from Gaia BHs, especially Gaia BH2. The visible star
of \bhb\ has subsolar metallicity and does not evolve to a red giant
star within the Hubble time, while Gaia BH2's visible star has solar
metallicity, and already evolves to a red giant star. However, if we
simulate the dynamical evolutions of many more open clusters, we are
likely to generate BH binaries similar to Gaia BH2.

We expect that Gaia BHs frequently have tertiary stars if they are
formed in open clusters. The frequency of the presence of tertiary
stars is highly uncertain. If we estimate the frequency from the small
samples, the frequency is $100$ \%, because the \bhb, which is just
one Gaia BH in our simulation, has a tertiary star. Even if we include
the BH binary with the secondary mass of $\sim 2 \msun$, this
frequency is not changed much because of its short lifetime. Even if
we take into account more many runs in Appendix
\ref{sec:EffectsOfInitialBinaryFraction}, the frequency is still high
$\sim 50$ \%. Nevertheless, we recognize that this is a rough
estimate. In order to determine the frequency accurately, we need to
perform many more cluster simulations, and acquire a larger number of
samples than now. On the other hand, there are no reports for such
tertiary stars in Gaia BHs. Its reason may be just that no one
searches for such tertiary stars \citep[but
  see][]{2021MNRAS.506.2269E}. Moreover, the tertiary star of the
\bhb\ becomes a white dwarf after $\sim 2$ Gyr, and may be too faint
to be detected by \gaia\ at the present day. Thus, we are not always
able to discover such tertiary stars if any.

As for multiplicity, our results show that a visible binary can orbit
a BH. In other words, the BH itself is a tertiary star of the multiple
star system. We observe several binaries orbiting BHs in our
simulations. However, their outer binary periods are $>10^5$
days. They cannot be discovered by \gaia, since their periods are much
larger than \gaia's operation duration, a few $10^3$ days. We may
generate such systems detectable by \gaia\ if we increase the number
of open cluster simulations again.

It appears that there are large uncertainties in our results, in other
words, the formation efficiency of Gaia BHs and multiplicity
frequency. This is derived from only one Gaia BHs formed from open
clusters with $3 \times 10^5 \msun$ in total. However, we expect that
our results would be robust. This is because our results are not
different from those in Appendix
\ref{sec:EffectsOfInitialBinaryFraction} by an order of magnitude. In
Appendix \ref{sec:EffectsOfInitialBinaryFraction}, we investigate open
clusters with $8 \times 10^6 \msun$ in total and the number of Gaia
BH-like binaries is $11$.

At the moment when we posted this paper to arXiv, only
\cite{2020PASJ...72...45S} have previously investigated Gaia BHs
formed in open clusters, although many studies have studied Gaia BHs
formed in isolated binaries \citep{2017MNRAS.470.2611M,
  2017ApJ...850L..13B, 2018ApJ...861...21Y, 2018arXiv181009721K,
  2018MNRAS.481..930Y, 2019ApJ...885..151S, 2019ApJ...886...68A,
  2020ApJ...905..134W, 2022ApJ...931..107C, 2022ApJ...928...13S,
  2023arXiv230107207S}. They have not found BH binaries with $\sim 1
\msun$ MS stars despite that they treat open clusters with $9 \times
10^5 \msun$ in total, $3$ times larger than those in this \paper.  No
Gaia BH-like binary in \cite{2020PASJ...72...45S} can be explained by
small statistics. No Gaia BH-like binary can be formed even in some of
model grids in which open clusters have $10^6 \msun$ in total
according to Appendix \ref{sec:EffectsOfInitialBinaryFraction}
regardless of binary fractions.

We should make a caveat that the presence of tertiary stars of Gaia
BHs does not mean the evidence of Gaia BH formation in open
clusters. A few 10 \% of galactic field stars are in triple and
quadruple systems \citep{2014AJ....147...87T,
  2017ApJS..230...15M}. Such multiple systems may form Gaia BHs with
tertiary stars. We expect that this possibility will be investigated
owing to rapid progress for studying evolution of multiple systems
\citep{2021MNRAS.502.4479H, 2022A&A...661A..61T, 2022MNRAS.516.1406S,
  2023arXiv230704793D}, although it has not
yet.\footnote{\cite{2023MNRAS.518.1057E, 2023MNRAS.521.4323E} form
Gaia BHs from triple systems. However these Gaia BHs do not have
tertiary stars, since the inner binaries merge in the halfway of their
evolutions.} Comparison between tertiary stars of Gaia BHs formed in
open clusters and multiple systems will helpful for identifying the
origin of Gaia BHs. In any case, what we can say now is that the open
cluster scenario will be ruled out if no Gaia BHs have tertiary stars
at all.

In the near future, we will perform a larger number of open cluster
simulations than in this \paper. These open clusters will have
different mass and density, and will be in different tidal fields. It
will reveal the formation efficiency of Gaia BHs in open clusters more
accurately. We will analyze the distribution of BH and visible star
masses, binary periods, eccentricities, and frequency of the presence
of tertiary stars. It is also important to obtain the distribution of
the tertiary star masses, and outer binary periods and
eccentricities. If an outer binary has a short enough period to cause
the ZLK oscillation of its inner BH binary, we may observe an
eccentricity variability of the BH binary \citep{2020ApJ...897...29H}.
We may also see visible binaries orbiting BHs with periods of
$100$-$2000$ days. The presence of such binaries may be indicated by
differences between spectroscopic and astrometric mass functions for
each binary \citep{2023ApJ...946...79T}. These distributions should be
a clue to elucidate the formation process of Gaia BHs.

This \paper\ indicates two important possibilities. First, we do not
need to rebuild the theory of common envelope evolution, which is
carefully constructed to explain several types of compact binaries,
such as double BHs \citep[e.g.][]{2016Natur.534..512B}. Second, if
Gaia BHs are dominantly formed in open clusters, they can be a useful
probe to study open cluster dynamics.

\section*{Acknowledgments}

This research could not be accomplished without the support by
Grants-in-Aid for Scientific Research (17H06360, 19K03907) from the
Japan Society for the Promotion of Science.  M.F. is supported by The
University of Tokyo Excellent Young Researcher Program. S.C. is
supported by the Fulbright U.S. Student Program, which is sponsored by
the U.S. Department of State and Japan-U.S.  Educational
Commission. Its contents are solely the responsibility of the author
and do not necessarily represent the official views of the Fulbright
Program, the Government of the United States, or the Japan-U.S.
Educational Commission. M.S. thanks a support by Research Fellowships
of Japan Society for the Promotion of Science for Young Scientists, by
Forefront Physics and Mathematics Program to Drive Transformation
(FoPM), a World-leading Innovative Graduate Study (WINGS) Program, the
University of Tokyo, and by JSPS Overseas Challenge Program for Young
Researchers. L.W. thanks the support from the one-hundred-talent
project of Sun Yat-sen University, the Fundamental Research Funds for
the Central Universities, Sun Yat-sen University (22hytd09) and the
National Natural Science Foundation of China through grant 12073090
and 12233013. Numerical simulations are carried out on Small Parallel
Computers at Center for Computational Astrophysics, National
Astronomical Observatory of Japan, the Yukawa Institute Computer
Facility, and Cygnus/Pegasus at the CCS, University of Tsukuba.

\section*{Data availability}

Results will be shared on reasonable request to authors. The data is
generated by the software {\tt PeTar} and {\tt SDAR}, which are
available in GitHub at \url{https://github.com/lwang-astro/PeTar} and
\url{https://github.com/lwang-astro/SDAR}, respectively. The initial
conditions of star cluster models are generated by the software {\tt
  MCLUSTER} \citep{2011MNRAS.417.2300K}, which is available in GitHub
at \url{https://github.com/lwang-astro/mcluster}.

\appendix

\section{Comparison between PeTar and NBODY6++GPU}
\label{sec:ComparisonBetweenPetarAndNbody6}

We make comparison between simulation results of {\tt PeTar} and {\tt
  NBODY6++GPU} \citep{2012MNRAS.424..545N, 2015MNRAS.450.4070W,
  2022MNRAS.511.4060K}. Indeed, \cite{2020MNRAS.497..536W} have
already done in detail. However, they have treated small globular
clusters with $\sim 10^5$ $\msun$, while we treat open clusters with
$\sim 10^3$ $\msun$. Thus, it might be worth comparing simulation
results between {\tt PeTar} and {\tt NBODY6++GPU} with respect of
cluster mass scale of $\sim 10^3$ $\msun$.

We generate 10 open clusters for each code. The metallicity is
$Z=0.02$, and initial binary fractions are 20 \%. We do not take into
account galactic tides, since {\tt NBODY6++GPU} does not support
combination with {\tt GALPY}. The other setup is the same as section
\ref{sec:Method}. Note that we do not remove any stars from
simulations even when stars go away from open clusters. This is true
for simulations in section \ref{sec:Method}.

\begin{figure}
  \includegraphics[width=\columnwidth]{\fdir/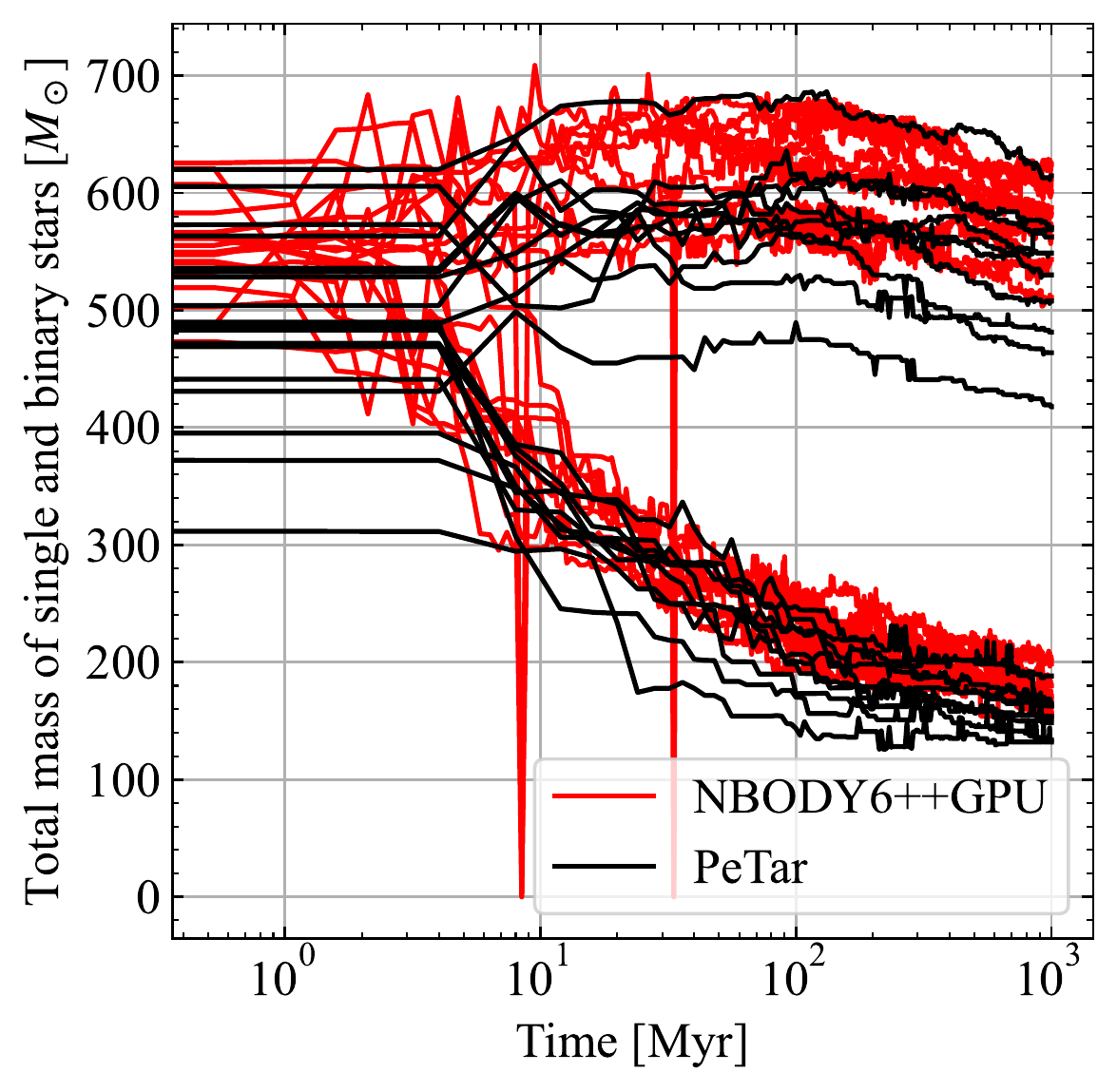}
  \includegraphics[width=\columnwidth]{\fdir/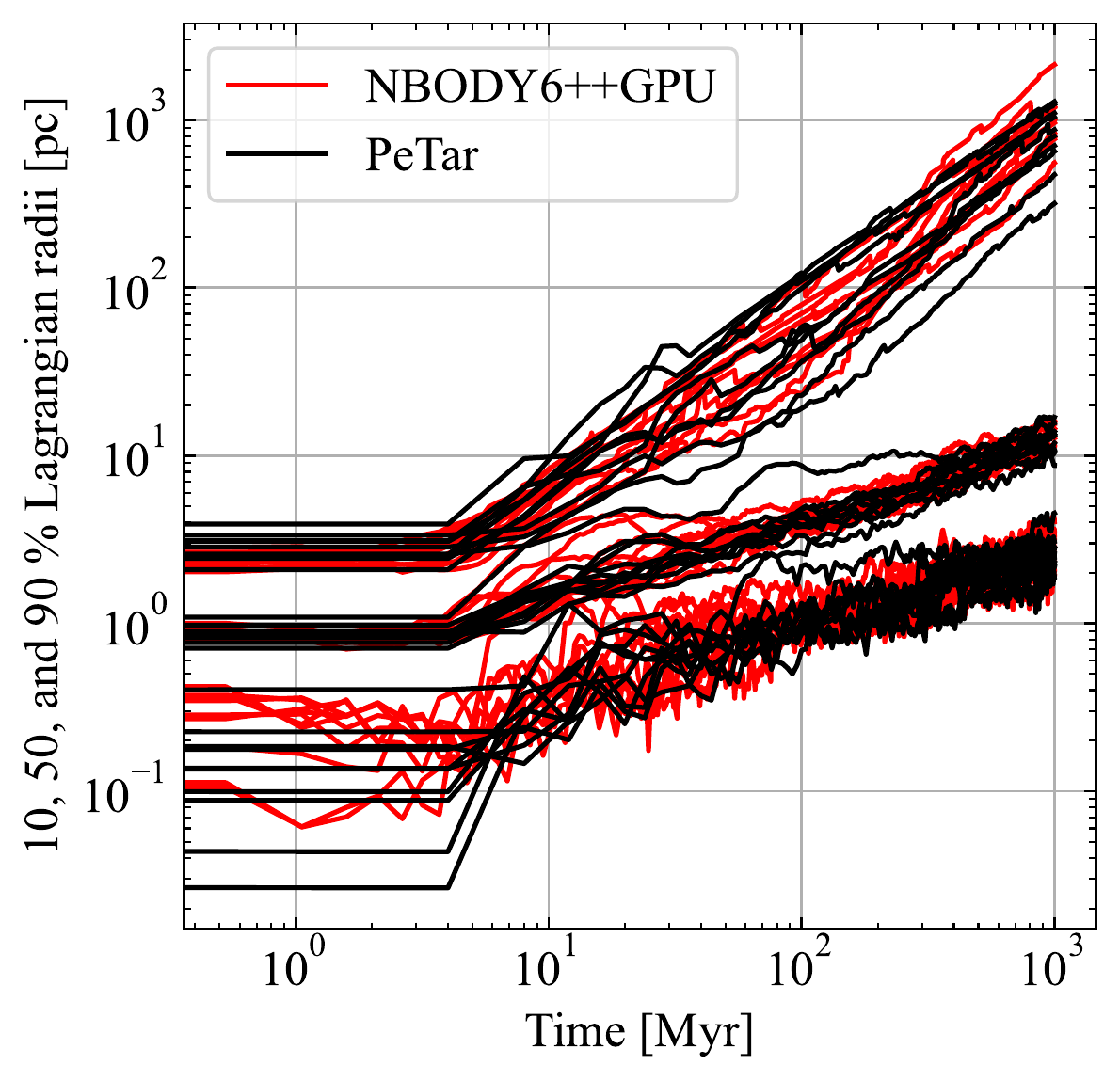}
  \caption{(Top) Time evolution of the total masses of single and
    binary stars. (Bottom) Time evolution of 10, 50, and 90 \%
    Lagrangian radii.}
  \label{fig:nbody6_time}
\end{figure}

\begin{figure}
  \includegraphics[width=\columnwidth]{\fdir/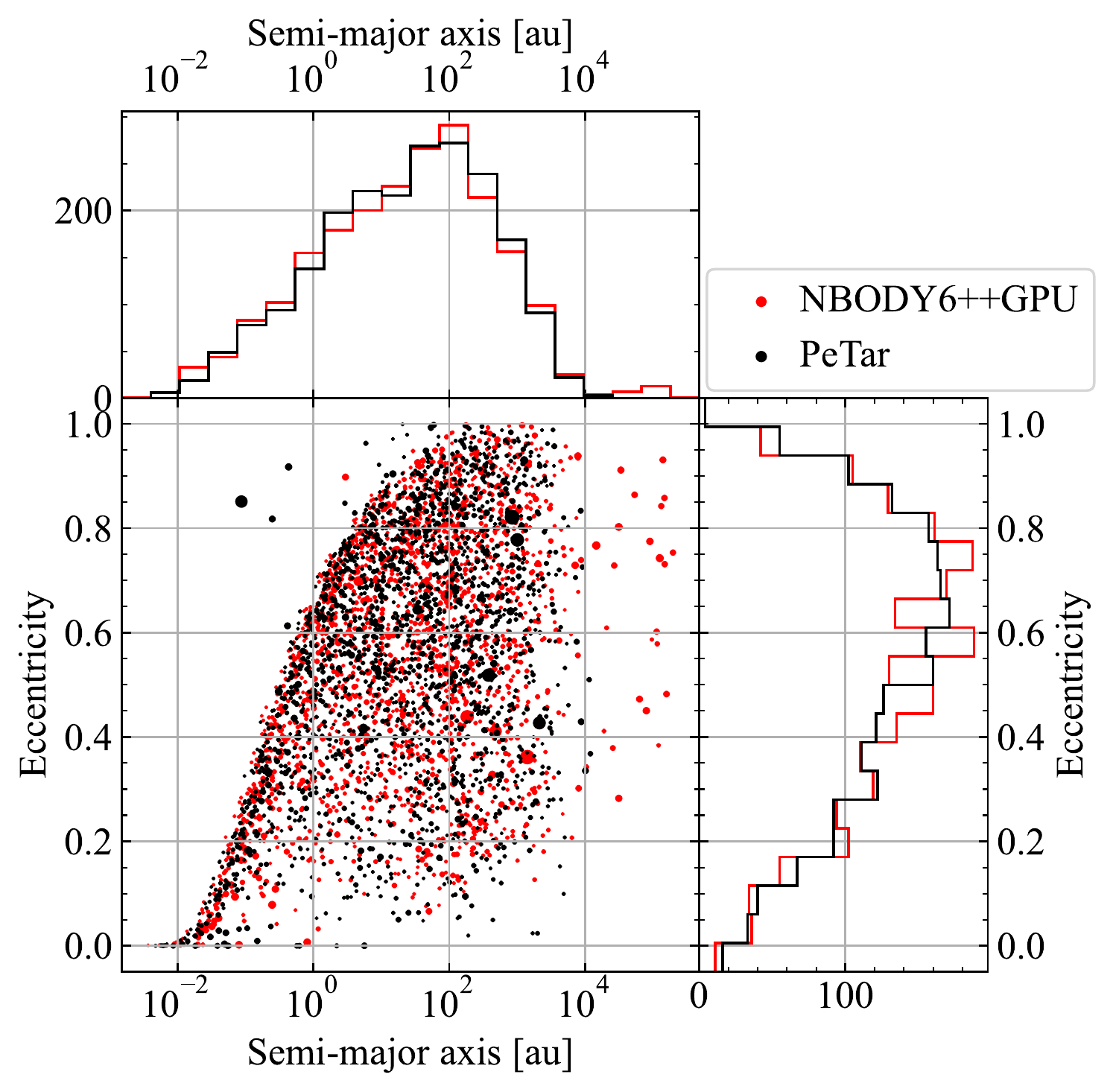}
  \includegraphics[width=\columnwidth]{\fdir/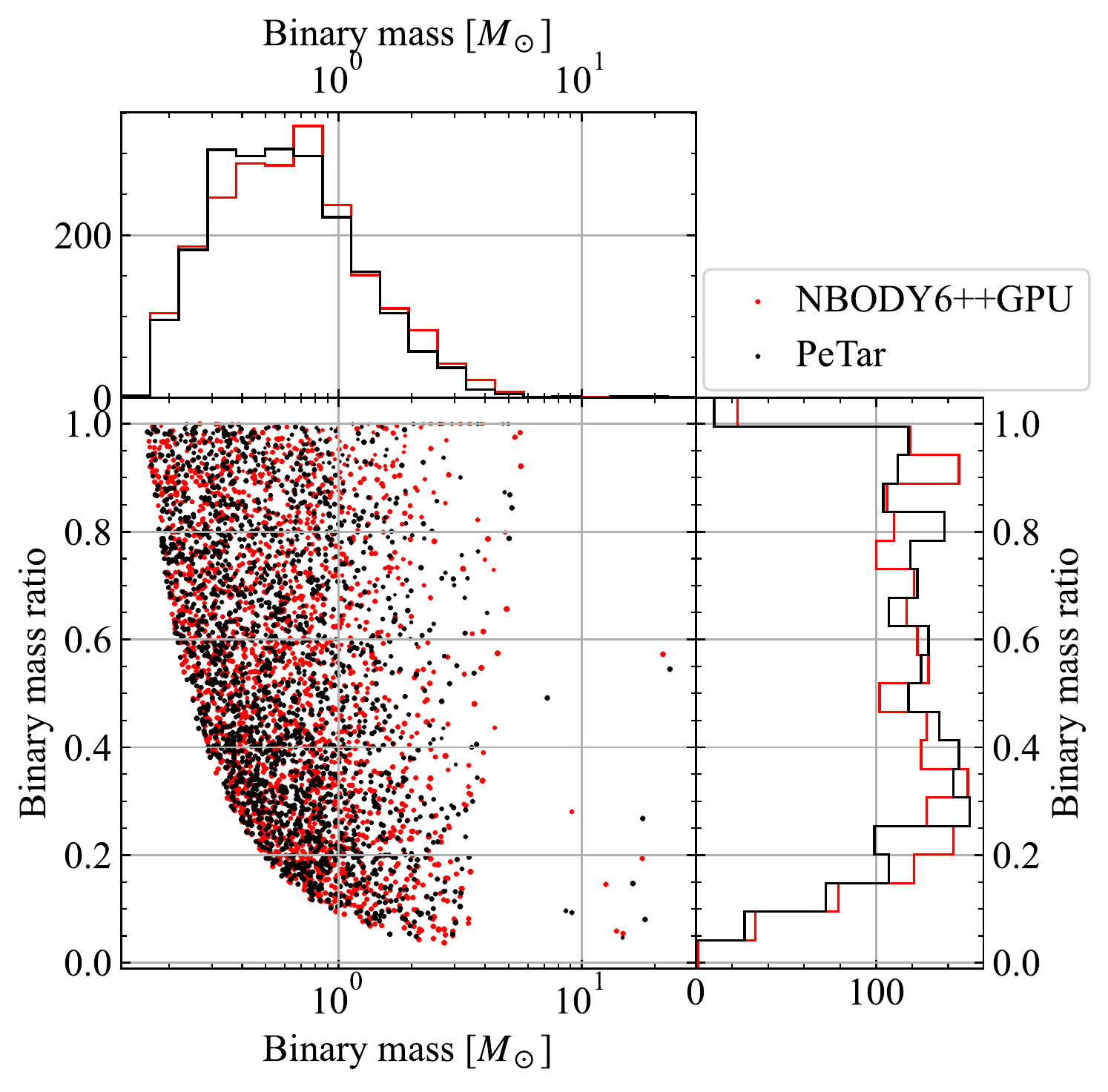}
  \caption{(Top) Distribution of binary semi-major axes and
    eccentricity at 500 Myr. The point sizes show binary masses on
    linear scale. (Bottom) Distribution of binary mass and mass ratio
    at 500 Myr. The point size indicates binary semi-major axes on
    logarithmic scale.}
  \label{fig:nbody6_binary}
\end{figure}

Figures \ref{fig:nbody6_time} and \ref{fig:nbody6_binary} shows the
cluster evolution and binary properties at 500 Myr. We do not find any
different features between {\tt PeTar} and {\tt NBODY6++GPU}
results. This indicates that {\tt PeTar} works well even for mass
scale of open clusters.

\section{Effects of initial binary fractions}
\label{sec:EffectsOfInitialBinaryFraction}

We investigate how the formation efficiency of Gaia BHs depend on
initial binary fractions. We adopt the fractions of 0, 20, 50, and 100
\% for open clusters with $Z=0.02$ and $0.005$. We generate 1000 open
clusters for each fraction and each metallicity in order to reduce
statistical uncertainties.

\begin{figure}
  \includegraphics[width=\columnwidth]{\fdir/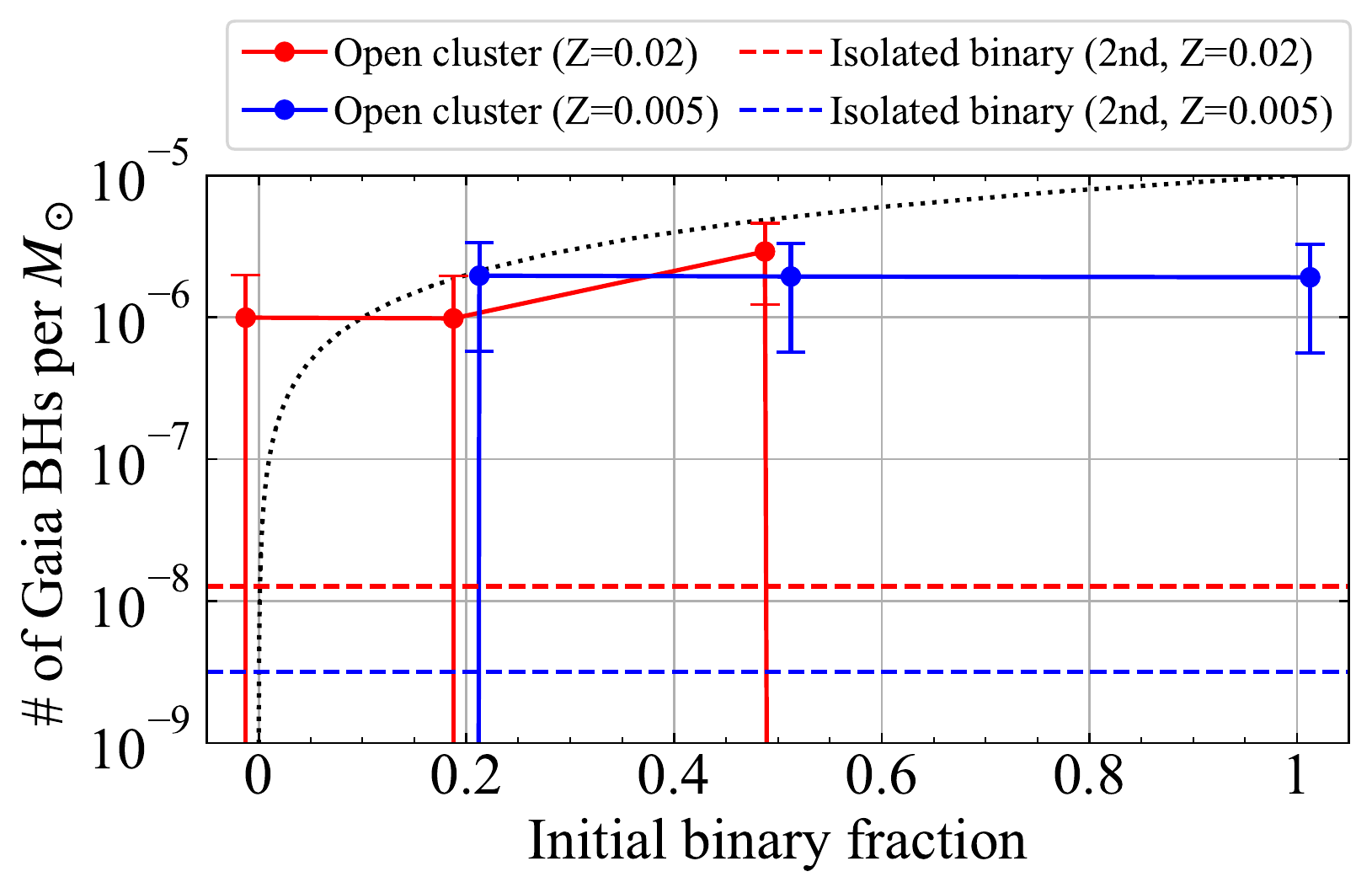}
  \includegraphics[width=\columnwidth]{\fdir/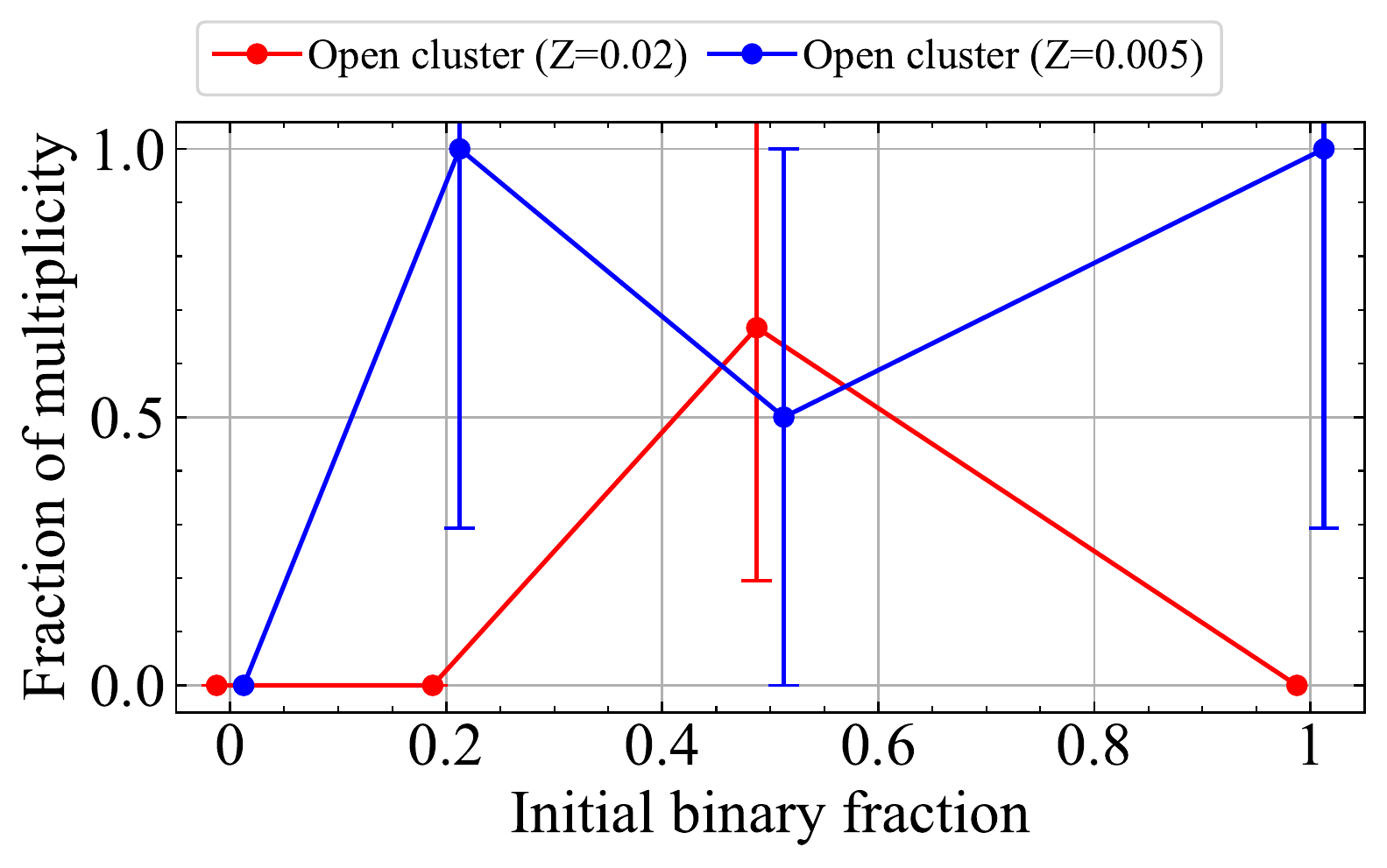}
  \caption{(Top) Formation efficiency of Gaia BHs in open clusters
    (solid curves) as a function of initial binary fractions. Error
    bars indicate standard errors under assumption that Gaia BH
    formation is a Poisson process. Dashed lines show the formation
    efficiency in isolated binaries. In order to be easy to see, the
    initial binary fractions are shifted by $0.125$ leftward and
    rightward for $Z=0.02$ and $0.05$, respectively. The definition of
    Gaia BHs is the same as section \ref{sec:Method}. The black dotted
    curve indicates the formation efficiency proportional to the
    initial binary fraction. (Bottom) Fraction of Gaia BHs in multiple
    star systems as a function of initial binary fractions. We
    calculate standard errors, assuming that Gaia BHs obtain tertiary
    stars in a Poisson process. The initial binary fractions are
    shifted in the same way as the top panel.}
  \label{fig:fbinary}
\end{figure}

The top panel of Figure \ref{fig:fbinary} shows that the formation
efficiency of Gaia BHs in open clusters is not sensitive to initial
binary fractions in open clusters. Although the error bars are large,
the average values are not dependent on the initial binary
fractions. The bottom panel of Figure \ref{fig:fbinary} indicates
little dependence of multiplicity on initial binary
fractions. Although the results suffer from small statistics, the
multiplicity does not monotonically increase with initial binary
fractions. These results mean that our choice of initial binary
fractions does not affect our results in section \ref{sec:Results}.

The top panel of Figure \ref{fig:fbinary} shows $\eta \sim 2 \times
10^{-6} M_\odot^{-1}$ for all binary fractions and
metallicities. Then, we can rewrite Equation (\ref{eq:GaiaBH1}) as
\begin{align}
  N_{\rm Gaia BH,MW} &\sim 1.2 \times 10^4 \left( \frac{\eta}{2 \times
    10^{-6} \msun} \right) \left( \frac{M_{\rm MW}}{6.1 \times 10^{10}
    \msun} \right) \nonumber \\
  &\times \left( \frac{f_{\rm cluster}}{0.1}
  \right). \label{eq:GaiaBH2}
\end{align}
This is similar to the estimate in Equation (\ref{eq:GaiaBH1}).

\begin{figure}
  \includegraphics[width=\columnwidth]{\fdir/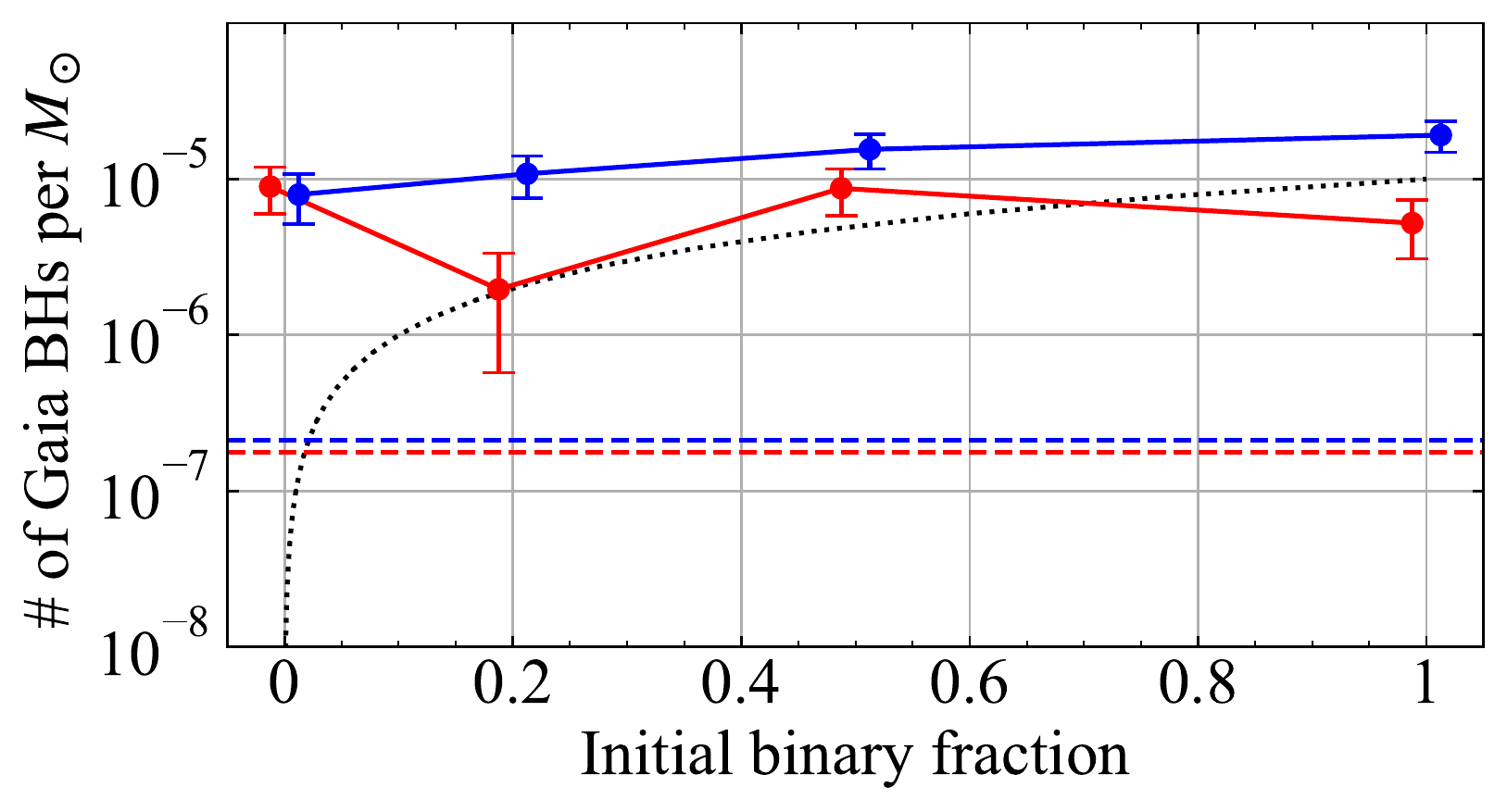}
  \includegraphics[width=\columnwidth]{\fdir/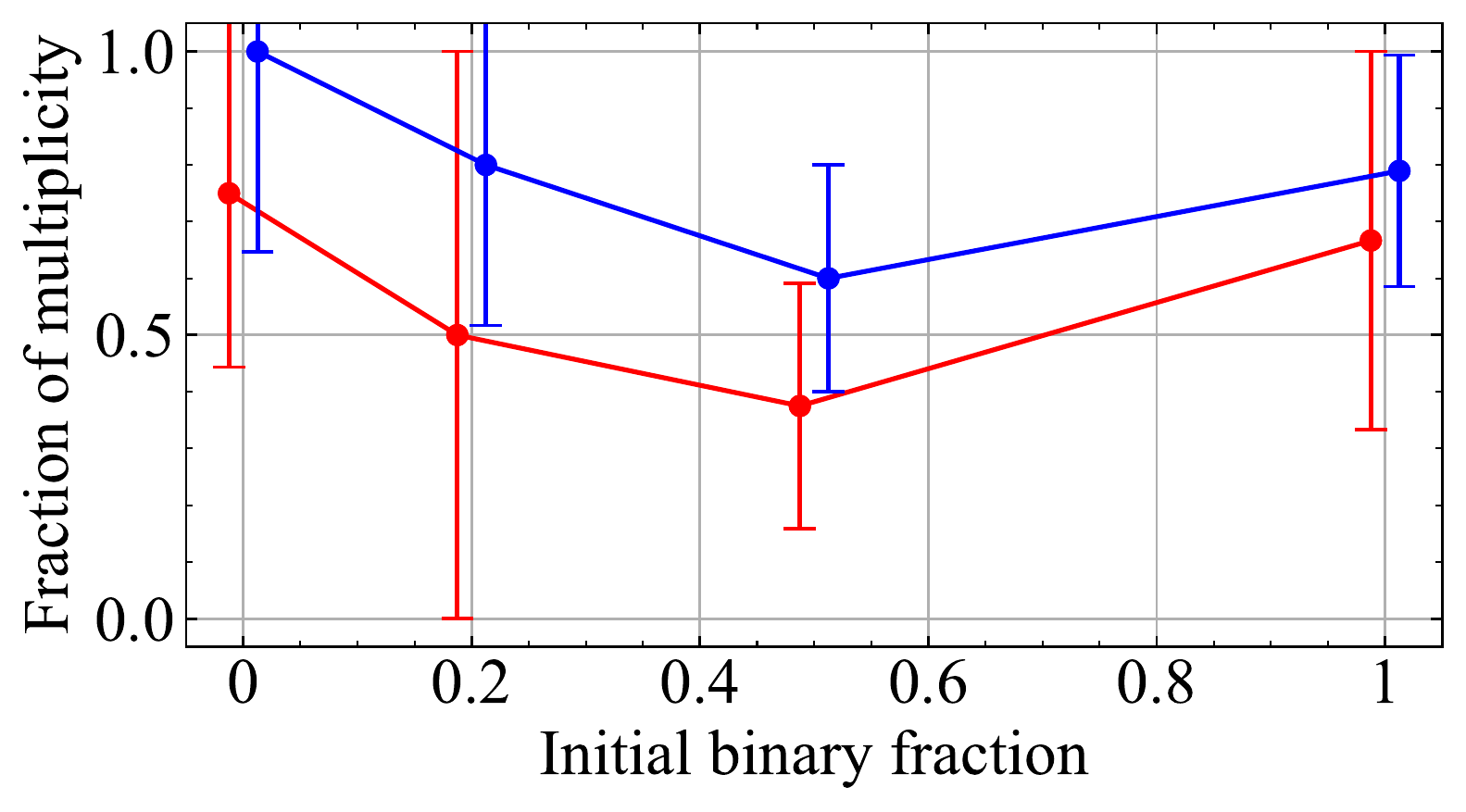}
  \caption{The same as Figure \ref{fig:fbinary}, except for the
    definition of Gaia BHs. Period and eccentricity criteria are
    relaxed, such that the former and latter are $0-4000$ days, and
    $0.0-1.0$, respectively. The companion criteria become more
    strict, such that $<0.8M_\odot$. Note that $<0.8M_\odot$ stars do
    not evolve to giant stars within the Hubble time.}
  \label{fig:fbinary}
\end{figure}

We show the formation efficiency and multiplicity fraction of Gaia
BHs. Here, we change the definition of Gaia BHs to mimic the
definition in \cite{2023arXiv230613121N}, such that period is $0-4000$
days, eccentricity is $0.0-1.0$, and the companion mass is
$<0.8M_\odot$. Compared to the definition in section \ref{sec:Method},
the formation efficiency is increased to $\sim 10^{-5}M_\odot^{-1}$ by
10 times. This is in good agreement with the results of
\cite{2023arXiv230613121N}. The multiplicity of Gaia BHs in open
clusters is $\sim 0.5$. As seen in the top panel, the formation
efficiency may increase with metallicity decreasing. However,
metallicity dependence is beyond the scope of this paper.

In summary, the formation efficiency of Gaia BHs is quite consistent
with the results of \cite{2023arXiv230614679R} and
\cite{2023arXiv230613121N}. Our formation efficiency looks different
from the formation efficiency of \cite{2023arXiv230613121N}. However,
the difference comes from the difference of the definitions of Gaia
BHs.

\bibliographystyle{mnras}

\bsp	% typesetting comment
\label{lastpage}
\end{document}